\documentclass{article}
\usepackage{spconf,amsmath,amssymb,graphicx}
\usepackage{color}

\usepackage{url}
\usepackage[subrefformat=parens]{subcaption}
\title{NEURAL NETWORK-BASED VIRTUAL MICROPHONE ESTIMATION WITH\\VIRTUAL MICROPHONE AND BEAMFORMER-LEVEL MULTI-TASK LOSS}
\name{\shortstack{Hanako Segawa$^1$, Tsubasa Ochiai$^2$, Marc Delcroix$^2$, Tomohiro Nakatani$^2$,\\\textit{Rintaro Ikeshita}$^2$, \textit{Shoko Araki}$^2$, \textit{Takeshi Yamada}$^1$, \textit{Shoji Makino}$^{1,3}$}}
\address{
  $^1$University of Tsukuba, Japan,
  $^2$NTT corporation, Japan,
  $^3$Waseda University, Japan}
\begin{document}
\ninept
\maketitle
\begin{abstract}

Array processing performance depends on the number of microphones available. Virtual microphone estimation (VME) has been proposed to increase the number of microphone signals artificially. Neural network-based VME (NN-VME) trains an NN with a VM-level loss to predict a signal at a microphone location that is available during training but not at inference. However, this training objective may not be optimal for a specific array processing back-end, such as beamforming. An alternative approach is to use a training objective considering the array-processing back-end, such as a loss on the beamformer output. This approach may generate signals optimal for beamforming but not physically grounded. To combine the advantages of both approaches, this paper proposes a multi-task loss for NN-VME that combines both VM-level and beamformer-level losses. We evaluate the proposed multi-task NN-VME on multi-talker underdetermined conditions and show that it achieves a 33.1 \% relative WER improvement compared to using only real microphones and 10.8 \% compared to using a prior NN-VME approach.

\end{abstract}
\section{Introduction}

Array signal processing \cite{benesty2008microphone, van1988beamforming,brandstein2001microphone} utilizing spatial information captured with multiple microphones has been actively studied for several decades. It plays a key role in developing various audio processing applications such as noise reduction, source separation, and source localization.
However, the achievable performance of an array signal-processing back-end, such as beamforming (BF), relies on the number of available microphones.
For example, BF with $C$ microphones enhances (separates) a specific sound source by producing $C-1$ nulls to reduce the interference sources. Accordingly, it cannot suppress all of the interference sources if the number of sound sources $I$ exceeds the number of microphones $C$, i.e., underdetermined condition ($C < I$).

To mitigate such performance limitations due to the number of available microphones, the virtual microphone estimation (VME) approach has been studied \cite{katahira2016nonlinear,yamaoka2019cnn,ochiai2021neural}. VME virtually increases the number of microphones by generating virtual observations at positions where there are no real microphones (virtual microphone, VM) given a few real observations (real microphone, RM).

Earlier studies \cite{katahira2016nonlinear,yamaoka2019cnn} have estimated the VM signals by linearly interpolating the phases of two RMs while relying on physical model-based assumptions: 1) plane wave propagation, 2) W-disjoint orthogonality of the sources \cite{yilmaz2004blind}, and 3) short inter-microphone distances to avoid spatial aliasing.
However, such assumptions may not always hold in realistic acoustic conditions, such as under reverberant and diffuse noise conditions.

A previous work \cite{ochiai2021neural} proposed a fully data-driven neural network-based VME framework (NN-VME) as an alternative VME approach that does not explicitly rely on the above assumptions.
NN-VME exploits the success of recent time-domain NN to estimate the waveform of the VM directly.
In other words, it can estimate both the amplitude and phase based on the supervised learning framework.
NN-VME does not make physical model-based assumptions but instead assumes that we can access RM observations at VMs' locations during the training stage, which is missing during the inference stage due to structural constraints and cost restrictions.
It trains the NN with a VM-level loss, which consists of minimizing the distance between the estimated VM and the RM observation at that location.
Consequently, it can estimate a signal that is close to the RM signal at the VM's location and virtually increases the number of microphones by augmenting the RMs with the estimated VMs.
This training scheme does not depend on a specific array signal processing back-end, such as BF.
Consequently, the VM signals generated by NN-VME could be applied to arbitrary array processing back-ends \cite{ochiai2021neural,segawa2022neural}.

Training an NN-VME with a VM-level loss offers versatility, but it may not be optimal for a specific array processing back-end.
If the array processing back-end is determined in advance, we could estimate VM signals better suited for that back-end by adopting a training loss on the output of the array processing back-end.
In this case, however, the estimated VM signal may not be interpretable as a virtually recorded observation signal.
For example, when using a BF-level training loss, the NN-VME may learn an extreme behavior, such as estimating only the target source signals, because the BF-level loss could still be improved even if the estimated VM signal is close to the target source signal.
Arguably, a signal trained with such an extreme tendency cannot be called a VM.

In this paper, we focus on the popular BF approach as the array processing back-end and consider a multi-talker scenario. We extend the NN-VME framework to adopt the BF-level loss by additionally assuming the availability of reference single-talker sources.
To make the NN-VME work for arbitrary positions of the target and interference sources, we propose combining the NN-VME with a mask-based frequency-domain BF \cite{heymann2016neural,erdogan2016improved} and the permutation invariant training (PIT) \cite{kolbaek2017multitalker} schemes.
Moreover, we propose a novel multi-task training objective for the NN-VME. It combines both VM-level and BF-level losses to take advantage of both training objectives by assuming the availability of both reference RM observations at the locations of the VMs and reference single-talker sources as the training targets.

We evaluate the effectiveness of the proposed multi-task NN-VME on the underdetermined multi-talker and reverberant acoustic conditions using three criteria: 1) estimation accuracy of the VM signal, 2) estimation accuracy of the beamformed signal augmented with the VME, and 3) speech recognition accuracy of the beamformed signal.
The experiments confirm that the NN-VME trained with the proposed multi-task loss successfully takes advantage of both loss functions and achieves higher performance than the NN-VMEs trained with the single-task losses.

\section{Related work}

Prior to this work, Yamaoka et al. \cite{yamaoka2019cnn} proposed the adoption of a BF-level training objective for the physical model-based VME framework, where the VM's amplitude is estimated using neural networks.
However, their experimental validation was relatively limited. For example, their experiments assume that the direction of the target source is known (i.e., the oracle steering vector of the target source is available) and that the positions of the target source and interference sources are fixed for all of the experiments.
Moreover, their experiments showed that such a BF-level training objective results in a large performance improvement for a closed (training) dataset but only a small improvement for an open (evaluation) dataset due to the severe nonlinear noise (artifact) in the estimated VM, probably as a result of the over-fitting \cite{yamaoka2019cnn}.
Therefore, the previous study \cite{yamaoka2019cnn} has not fully revealed the effectiveness of the BF-level training objective for the VME framework in a practical use case, i.e., the direction of the target source is unknown and variable.

This paper incorporates a BF-level training objective in a recent fully data-driven NN-VME framework \cite{ochiai2021neural,segawa2022neural} by combining 1) the mask-based frequency-domain BF \cite{heymann2016neural,erdogan2016improved} that does not use any prior information of the sources and 2) the PIT-based multi-talker reconstruction loss \cite{kolbaek2017multitalker}.
We experimentally confirmed that the proposed NN-VME with the BF-level loss successfully improves BF performance for an open evaluation dataset that contains variable source positions.

\section{Proposed method: Multi-task training loss for NN-VME}

In this paper, we assume a situation where $C$ microphones record $I$ source signals.
Let $\mathbf{s}_{i} \in \mathbb{R}^{T}$ denote a time-domain source signal of the $i$-th source, where $T$ is the length of the waveform (i.e., number of samples).
The observed signal $\mathbf{y}_{c} \in \mathbb{R}^{T}$ is modeled as $\mathbf{y}_{c} =  \sum_{i=1}^{I} \mathbf{x}_{c,i} + \mathbf{n}_{c}$, where $\mathbf{x}_{c,i}$ denotes a reverberant source signal (i.e., spatial image) of the $i$-th source recorded at the $c$-th channel, and $\mathbf{n}_{c} \in \mathbb{R}^{T}$ is the additive noise signal.

\vspace{-1mm}
\subsection{General procedure of NN-VME}

The VME framework consists of two steps: 1) estimating the VM and 2) applying the array signal processing technique.

In the NN-VME \cite{ochiai2021neural}, a time-domain signal estimation neural network \cite{luo2019conv} is used to estimate the amplitude and phase of the VM signal simultaneously.
Let $\mathbf{r}_{c} \in \mathbb{R}^{T}$ denote the time-domain observed signal recorded by the $c$-th RM, and $\widehat{\mathbf{v}}_{c'} \in \mathbb{R}^{T}$ denote the estimated signal corresponding to the $c'$-th VM.
Given the RM observation $\mathbf{r} = \{ \mathbf{r}_{c=1}, \ldots , \mathbf{r}_{c=C_r} \} \in \mathbb{R}^{T \times C_r}$, the VM observation $\widehat{\mathbf{v}} = \{ \widehat{\mathbf{v}}_{c'=1}, \ldots , \widehat{\mathbf{v}}_{c'=C_v} \} \in \mathbb{R}^{T \times C_v}$ is estimated as:
\begin{equation}
\label{Eq:nnvme}
    \widehat{\mathbf{v}}={\text{NN-VME}} (\mathbf{r}),
\end{equation}
where ${\text{NN-VME}} (\cdot)$ denotes the neural network model, and $C_r$ and $C_v$ denote the number of RMs used as input and the number of VMs to be estimated, respectively.

When using the VM with a microphone array processing back-end, the estimated VM signal is combined with the RM signal to obtain an augmented microphone array signal $\overline{\mathbf{y}} = [\mathbf{r},  \widehat{\mathbf{v}}] \in \mathbb{R}^{T \times C}$, where $C = C_{r}+C_{v}$.
We expect the array processing performance will improve when using augmented microphone array signal $\overline{\mathbf{y}}$, whose number of microphones is virtually increased, compared to using only the real array observation $\mathbf{r}$.

This paper focuses on the source separation task and adopts a mask-based frequency-domain BF \cite{heymann2016neural,erdogan2016improved} as the array processing back-end.
Given the augmented array observation $\overline{\mathbf{y}}$, the BF computes the enhanced signal $\widehat{\mathbf{x}}_{i}^{\text{BF}}\in \mathbb{R}^{T}$ of the $i$-th source as:
\begin{equation}
 \label{eq:bf}
     \widehat{\mathbf{x}}_{i}^{\text{BF}} = \text{MaskBF}_{i}(\overline{\mathbf{y}}),
 \end{equation}
where $\text{MaskBF}_{i}(\cdot)$ denotes the functional representation of the mask-based BF for the $i$-th source.
Specifically, in this paper, we adopt the formulation of the Minimum Variance Distortionless Response (MVDR) BF of \cite{souden2009optimal}.
We also follow prior studies \cite{ochiai2021neural,segawa2022neural} in implementing mask-based BF with NN-VME.

\subsection{Loss function of NN-VME}

In this section, we first overview two levels of loss functions (i.e., VM-level and BF-level) and then introduce the proposed multi-task loss function.
Note that the training objective determines the property of the estimated VM signals.
For example, when using the VM-level loss \cite{ochiai2021neural}, the VM signal mimics the RM observation that would be captured at the position of the VM.
Here, the RM observations at the location of the VM are assumed to be known in the training stage but missing in the inference stage.
When using the BF-level loss \cite{yamaoka2019cnn}, the VM signal becomes a signal that improves the array processing performance when combined with the RM observations.

\subsubsection{Virtual microphone (VM)-level training loss}

The original NN-VME \cite{ochiai2021neural} adopts the VM-level training loss that brings the estimated VM signals $\widehat{\mathbf{v}}$ close to the target signals $\mathbf{v}$, i.e., RM observations at the locations of the VMs.

The NN-VME framework assumes that we have fewer constraints on the number of microphones during system development (i.e., training stage) than during actual deployment (i.e., inference stage).
It divides the observed signal $\mathbf{y} = \{ \mathbf{y}_{c=1}, \ldots , \mathbf{y}_{c=C} \} \in \mathbb{R}^{T \times C}$ into two subsets: one for network input (i.e., $\mathbf{r} \in \mathbb{R}^{T \times C_r}$) and the other for network target (i.e., $\mathbf{v} \in \mathbb{R}^{T \times C_v}$).
It assumes that a set of input and target signals $\{\mathbf{r}, \mathbf{v}\}$ is available for training the model, while only $\mathbf{r}$ is available for the inference stage, i.e., $\mathbf{v}$ is missing due to structural constraints or cost restrictions.
Given the set of input and target signals $\{\mathbf{r}, \mathbf{v}\}$, the VM-level loss is defined based on the classical signal-to-noise ratio (SNR) loss as:
\begin{align}
    \mathcal{L}_{\text{VM}} = \text{SNR}(\mathbf{v}, \widehat{\mathbf{v}}),
    \label{eq:L_VM}
\end{align}
where $\widehat{\mathbf{v}}$ is the output of the $\text{NN-VME}(\cdot)$ module, as in Eq.~\eqref{Eq:nnvme}.
Here, we adopt the widely used classical SNR \cite{le2019sdr} as a loss metric.
Given the time-domain reference signal $\mathbf{z}_{\text{ref}} \in \mathbb{R}^{T}$ and estimated signal $\mathbf{z}_{\text{est}} \in \mathbb{R}^{T}$, the SNR loss is computed as $\text{SNR}({\mathbf{z}}_{\text{ref}}, {\mathbf{z}}_{\text{est}}) = - 10 \log_{10} (||{{\mathbf{z}}_{\text{ref}}}||^2 / ||{\mathbf{z}}_{\text{ref}} - { \mathbf{z}}_{\text{est}}||^2 )$.

\subsubsection{Beamformer (BF)-level training loss}

In addition to the above VM-level loss function, we can consider the BF-level training loss that makes the estimated beamformed signals $\widehat{\mathbf{x}}_{i}^{\text{BF}}$ close to the target signals $\mathbf{x}_{c_{\text{ref}},i}$, i.e., the single-talker reverberant source.

In the supervised source separation, we assume that a set of input and target signals $\{\mathbf{r}, \mathbf{x}\}$ is available for training the model, where $\mathbf{x} = [ \mathbf{x}_{c_{\text{ref}},i=1}, \ldots \mathbf{x}_{c_{\text{ref}},i=I} ] \in \mathbb{R}^{T \times I}$ denotes the single-talker reverberant signals for each source at the reference channel $c_{\text{ref}}$.
Given the set of input and target signals $\{\mathbf{r}, \mathbf{x}\}$, the BF-level loss is defined based on the classical SNR loss as:
%
\begin{align}
    \mathcal{L}_{\text{BF}} = \min_{p \in \text{perm}(I)} \sum_{i=1}^{I} \text{SNR}(\mathbf{x}_{c_{\text{ref}},i}, \widehat{\mathbf{x}}_{p_i}^{\text{BF}}),
    \label{eq:L_BF}
\end{align}
where $\widehat{\mathbf{x}}_{p_i}^{\text{BF}}$ is the output of the $\text{MaskBF}_{i}(\cdot)$ module constructed from the augmented array observation $\overline{\mathbf{y}}$ with the NN-VME, as in Eq.~\eqref{eq:bf}.
$\text{perm}(I)$ produces all possible permutations, and $p \colon \{ 1, \ldots, I \} \to \{ 1, \ldots, I \}$
is a permutation that maps $i$ to $p_i \in \{ 1, \ldots, I \}$.
Here, we adopt the PIT scheme \cite{kolbaek2017multitalker} to handle the multiple sources.
We can expect that adopting the reconstruction loss for all sources would impose the constraint that the estimated VM signals maintain the spectral and spatial information of all sources, unlike \cite{yamaoka2019cnn}.

\subsubsection{Proposed multi-task training loss combining VM-level and BF-level losses}
\begin{figure}[tb]
\centering
 \includegraphics[width=0.95\columnwidth]{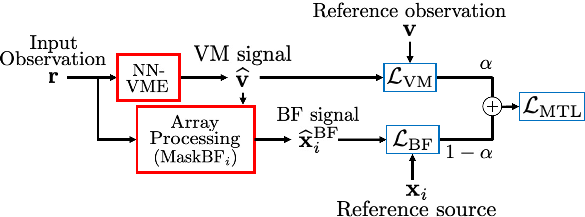}
\vspace{3mm}     
\caption{Multi-task learning with VM-level and BF-level losses.}
\label{fig:NN-VME}
\vspace{-10pt}   
\end{figure}
Figure~\ref{fig:NN-VME} shows a schematic diagram of the proposed multi-task learning scheme.
To take advantage of the two different levels of training objectives, we introduce a multi-task loss function $\mathcal{L}_{\text{MTL}}$, which combines the VM-level $\mathcal{L}_{\text{VM}}$ and BF-level $\mathcal{L}_{\text{BF}}$ loss functions:
\begin{align}
\label{eq:multi-loss}
    \mathcal{L}_{\text{MTL}} = \alpha\mathcal{L}_{\text{VM}} + (1-\alpha)\mathcal{L}_{\text{BF}},
\end{align}
where $0 \leq \alpha \leq 1$ represents the interpolation weight hyperparameter that controls the trade-off between VM-level and BF-level losses.

By combining the VM-level and BF-level losses, we expect the multi-task loss to suppress the over-fitting issues reported when using only BF-level loss \cite{yamaoka2019cnn} and to improve the source separation performance compared to using only the VM-level loss.
Moreover, it could maintain the properties of the VM, i.e., obtaining an estimated signal close to the signal that would be captured by an RM at that position.
Finally, we may also benefit from improved generalization thanks to the multi-task learning effect \cite{caruana1998multitask}.

\section{Experiment}

\subsection{Experimental conditions}
In this experiment, all of the training and evaluation data consisted of simulated reverberant noisy three-speaker mixtures using speech from the Wall Street Journal (WSJ) corpus \cite{paul1992design} and noise from the CHiME-3 corpus \cite{barker2015third}.
The room impulse responses were generated using the image method \cite{allen1979image}.
The reverberation time ($T_{60}$) was randomly selected from 0 ms to 300 ms for both training and evaluation data.
For each mixture, we randomly sampled the position of the speakers and the microphone array.
We simulated square rooms with width and depth randomly set to 2.5 $\sim$ 10 m, and the height set to 2.5 $\sim$ 5 m.
The signal-to-interference ratio (SIR) for interfering speakers was randomly set within the range of $-$3 dB $\sim$ 3 dB with respect to the first speaker, and the signal-to-noise ratio (SNR) of the diffuse noise was set to 20 dB.
We generated 30,000, 5,000, and 5,000 mixtures for the training, development, and evaluation sets, respectively.
The microphone geometry is a rectangular microphone array with six channels corresponding to the CHiME-3's tablet device \cite{barker2015third} (refer to the figure in \cite{barker2015third} for details).
In the following experiments, we used the three bottom channels, i.e., channels 4, 5, and 6, linearly arranged at 10 cm intervals, and assumed that channels 4 and 6 are the RMs and channel 5 is the VM.

\subsection{Evaluation systems}

The network architecture of the NN-VME was based on the time-domain convolutional network (TDCN) \cite{luo2019conv} as in our prior work \cite{ochiai2021neural,segawa2022neural}.
According to the notation of a previous study \cite{luo2019conv}, the hyperparameters are set to $N = 256$,
$L = 20$, $B = 256$, $H = 512$, $P = 3$, $X = 8$, and $R = 4$.
For the optimization, we used the Adam algorithm \cite{kingma2015adam} and gradient clipping \cite{pascanu2013difficulty} with an initial learning rate of $0.0001$, and we stopped the training procedure after 100 epochs.

We also prepared a TDCN-based source separation model (i.e., convolutional
time-domain audio separation network (Conv-TasNet) \cite{luo2019conv}) based on the PIT scheme \cite{kolbaek2017multitalker} to estimate the time-frequency masks for each source required to construct the mask-based BF \cite{heymann2016neural,erdogan2016improved}.
Following a previous study \cite{ochiai2020beam}, the time-frequency mask for each source is computed by applying STFT to the time-domain observed signal and the separated signal of each source and then taking the ratio of the magnitudes between them.

The network and optimization configurations of the separation model were basically the same as those of NN-VME.
The difference is that NN-VME has a single output and estimates the speech mixture at the locations of the VM, while the separation model has multiple outputs and estimates the clean speech for each source.

In this paper, we separately trained the parameters of NN-VME and the source separation model.
We first trained the source separation model and then optimized only the parameters of the NN-VME while constructing the BF with the estimated time-frequency masks.

\subsection{Evaluation metrics}

To evaluate the accuracy of the estimated VMs and their effectiveness for the array processing, we adopted two types of signal-to-distortion ratio (SDR) and word error rate (WER) by following a previous study \cite{ochiai2021neural}.
Given an estimated signal $\mathbf{z}_{\text{est}} \in \mathbb{R}^{T}$ and a reference signal $\mathbf{z}_{\text{ref}} \in \mathbb{R}^{T}$, the SDR is defined as $\text{SDR}(\mathbf{z}_{\text{ref}}, \mathbf{z}_{\text{est}})= 10\log_{10} (||{\mathbf{z}_{\text{tgt}}}||^2 / ||\mathbf{z}_{\text{tgt}} - \mathbf{z}_{\text{est}}||^2)$, where $\mathbf{z}_{\text{tgt}}$ is computed by orthogonally projecting the estimated signal $\mathbf{z}_{\text{est}}$ onto the reference signal $\mathbf{z}_{\text{ref}}$ \cite{vincent2006performance}.

First, we evaluated the accuracy of the estimated VM signal using $\text{SDR}_{\text{VM}}=\text{SDR}(\mathbf{v},\widehat{\mathbf{v}})$, where $\mathbf{v} \in \mathbb{R}^{T}$ denotes the RM observation at the position of the VM as the reference signal, and $\widehat{\mathbf{v}} \in \mathbb{R}^{T}$ denotes the estimated VM observation.

In addition, we measured the source separation performance of the beamformed signals using $\text{SDR}_{\text{BF}}=\text{SDR}({\mathbf{x}}, \widehat{\mathbf{x}}^{\text{BF}})$, where $\mathbf{x} \in \mathbb{R}^{T}$ denotes the reference single-talker reverberant signal (i.e., spatial image), and $\widehat{\mathbf{x}}^{\text{BF}} \in \mathbb{R}^{T}$ denotes the beamformed signal.
In the evaluation, we used the fourth channel as the reference microphone.
The $\text{SDR}_{\text{BF}}$ was computed for each of the three speakers in the mixture and then averaged to obtain the total score, where the permutations between the estimates and references were determined based on the SIR score.

Finally, to evaluate the automatic speech recognition (ASR) performance of the beamformed signals, we built a deep neural network-hidden Markov model (DNN-HMM) hybrid ASR system \cite{bourlard2012connectionist,hinton2012deep} based on the Kaldi's CHiME-4 recipe \cite{povey2011kaldi}.
The acoustic model was trained with a lattice-free maximum mutual information framework \cite{povey2016purely}.
As training data, we used 1) noisy single-talker signals, 2) beamformed signals using three RMs, and 3) beamformed signals using two RMs and one VM estimated by the NN-VME ($\alpha=1.0$).
We used a trigram language model for decoding.

\begin{figure}[t]

\begin{minipage}[b]{\linewidth}
\centering
\includegraphics[width=0.95\columnwidth]{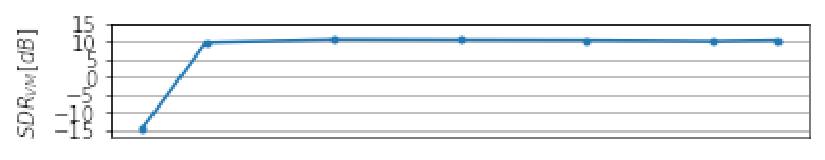}
\subcaption{$\text{SDR}_\text{VM}$}
\vspace{-0pt}
\label{fig:SDRVM}
\end{minipage}

\begin{minipage}[b]{\linewidth}
\centering
\includegraphics[width=0.95\columnwidth]{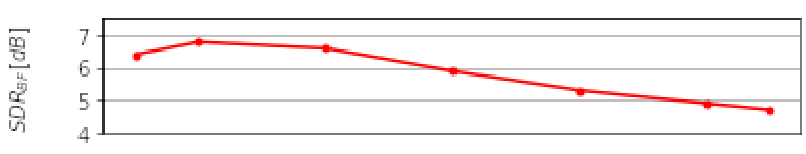}
\subcaption{$\text{SDR}_\text{BF}$}
\vspace{-0pt}
\label{fig:SDRAP}
\end{minipage}

\begin{minipage}[b]{\linewidth}
\centering
\includegraphics[width=0.95\columnwidth]{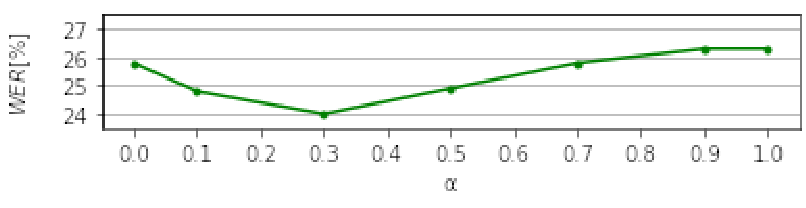}
\subcaption{WER}
\vspace{-0pt}
\label{fig:WER}
\end{minipage}

\caption{Impact of multi-task weight $\alpha$ on ${\text{SDR}_{\text{VM}}}$, ${\text{SDR}_{\text{BF}}}$ [dB] (higher is better) and WER [\%] (lower is better).}
\label{fig:alpha}
\vspace{-10pt}
\end{figure}

\subsection{Evaluation of multi-task interpolation weight}

First, we explore the impact of the multi-task interpolation weight $\alpha$ in Eq.~\eqref{eq:multi-loss} on the evaluation measures.
Figure~\ref{fig:alpha} shows $\text{SDR}_\text{VM}$, $\text{SDR}_\text{BF}$, and WER scores for the evaluated BF systems, where the x-axis indicates the multi-task interpolation weight $\alpha$ and the y-axis indicates each of the evaluation measures.
We varied the weight hyperparameter in the ranges of $\alpha = \{ 0.0, 0.1, 0.3, 0.5, 0.7, 0.9, 1.0 \}$.
Here, $\alpha=1.0$ and $\alpha=0.0$ correspond to the conventional single-task loss function; $\alpha=1.0$ is equivalent to using only the VM-level loss $\mathcal{L}_{\text{VM}}$ as in Eq.~\eqref{eq:L_VM}, while $\alpha=0.0$ is equivalent to using only the BF-level loss $\mathcal{L}_{\text{BF}}$ as in Eq.~\eqref{eq:L_BF}.

From the figure, we confirm that using the BF-level loss ($\alpha=0.0$) outperforms using only the VM-level loss ($\alpha=1.0$) in terms of $\text{SDR}_\text{BF}$ and WER. However, the VM signal is not related to the RM as shown by the low value of $\text{SDR}_\text{VM}$ for $\alpha=0.0$. By appropriately tuning $\alpha$, the proposed NN-VME with multi-task loss (e.g., $\alpha=0.3$) outperforms the single-task models, particularly in terms of WER.

\subsection{Overall evaluation}

Table~\ref{tbl:result} summarizes the signal-level 
and ASR-level 
performance measures for all the evaluated BFs.
Here, VM-BF denotes the BF constructed with two RMs (channels 4 and 6) and one VM (channel 5).
%
System (1) indicates the performance of the observed signal without processing.
Systems (2) and (3) denote the BF constructed with two (channels 4 and 6) and three (channels 4, 5, and 6) RMs, which would correspond to the lower-bound and upper-bound performances of the VM-BF.
Note that we are dealing with challenging conditions, i.e., the separation of noisy and reverberant three-speaker mixtures using a limited amount of microphones, which is reflected by the relatively low performance of the baseline system (3).
Systems (4) and (5) correspond to NN-VME trained with the BF-level and VM-level losses, respectively.
Systems (6) and (7) correspond to NN-VME trained with the proposed multi-task loss with two different weights.


The performance of the baseline system (2) is relatively low due to the underdetermined condition ($C < I$). Using the VM, we can virtually increase the number of microphones and make the systems determined, which explains the significant boost in the performance of systems (4) to (7).

Note that unlike the previous study \cite{yamaoka2019cnn}, our proposed NN-VME using PIT-based BF-level loss improves performance even for unseen conditions.
Moreover, the VM-BF trained with BF-level loss (i.e., system (5)) achieved better $\text{SDR}_\text{BF}$ and WER scores compared to the VM-BF trained with VM-level loss (i.e., system (4)), but the score of $\text{SDR}_\text{VM}$ becomes very low. 
This can be expected because using only the BF-level loss $\mathcal{L}_{\text{BF}}$ ($\alpha=0.0$) does not directly specify the property of the estimated VM signals, and thus there is no guarantee that the output of the NN-VME module imitates the observed RM signals that are actually recorded at the specific microphone position (i.e., channel 5 in this experiment).

On the other hand, systems (6) and (7) trained with the proposed multi-task loss retain high $\text{SDR}_\text{VM}$ scores.
In addition, probably due to the effect of the multi-task learning scheme \cite{caruana1998multitask}, they also achieved better $\text{SDR}_\text{BF}$ and WER scores than systems (4) and (5) trained with the single-task losses.
Here, the WER of system (7) with multi-task loss is significantly better than that of system (4) with single-task BF-level loss.
This is probably because the VM signal generated only with the BF-level loss becomes quite artificial and may result in containing more nonlinear distortions in the beamformed signals, which is reported to have a negative effect on ASR performance \cite{iwamoto2022how,zorilua2022speaker}.

These results show the effectiveness of optimizing the NN-VME while considering the specific array processing back-end.
Furthermore, they also show the effectiveness of using the multi-task training objective considering both the VM-level and BF-level losses, which enables the generation of interpretable VM signals and leads to better array processing performance.

\begin{table}[t]
\caption{${\text{SDR}_{\text{VM}}}$, ${\text{SDR}_{\text{BF}}}$ [dB] (higher is better) and WER [\%] (lower is better) for evaluated beamforming systems.}
\label{tbl:result}
\begin{center}
\vspace{-5mm}   
\scalebox{1.05}{
\begin{tabular}{@{} |c| c | c | c | c | c | @{}}
\hline
& Method & $\alpha$ & ${\text{SDR}_\text{VM}}$ & ${\text{SDR}_\text{BF}}$ & {WER}\\
\hline
(1)&Mixture & - & - & -3.1 & 98.4 \\
(2)&RM-BF (2ch) & - & - & 3.0 &  35.9\\
(3)&RM-BF (3ch) & - & - & 7.1 & 18.8\\
\hline
(4)&VM-BF ($\mathcal{L}_{\text{BF}}$) & 0.0 & -14.3 & 6.4 & 25.8 \\
(5)&VM-BF ($\mathcal{L}_{\text{VM}}$) & 1.0 & 10.5 & 4.7 & 26.9\\
\hline
(6)&VM-BF ($\mathcal{L}_{\text{MTL}}$) & 0.1 & 9.8 & $\boldsymbol{6.8}$ & 24.8 \\
(7)&VM-BF ($\mathcal{L}_{\text{MTL}}$) & 0.3 & $\boldsymbol{10.7}$ & 6.6 & $\boldsymbol{24.0}$ \\
\hline
\end{tabular}
}
\vspace{-7mm}   
\end{center}
\end{table}

\section{Conclusions}

This paper proposed a novel multi-task learning scheme for the NN-VME framework that combines the VM-level and BF-level losses.
We evaluated the effectiveness of the proposed method on the underdetermined source separation task with mask-based BF.
The experimental results show that the NN-VME with the proposed multi-task training loss achieved better source separation and speech recognition performances than the NN-VME with a single-task training loss, i.e., using only the VM-level loss or the BF-level loss.

\vfill\pagebreak

\bibliographystyle{IEEEbib}
\bibliography{mybib}

\end{document}